# Can the new Neutrino Telescopes reveal the Gravitational Properties of Antimatter?

Dragan Slavkov Hajdukovic[1]
PH Division CERN
CH-1211 Geneva 23;
dragan.hajdukovic@cern.ch
[1]On leave from Cetinje; Montenegro

**Abstract**
We argue that the hypothesis of the gravitational repulsion between matter and antimatter can be tested at the Ice Cube, a neutrino telescope, recently constructed at the South Pole. If there is such a gravitational repulsion, the gravitational field, deep inside the horizon of a black hole, might create neutrino-antineutrino pairs from the quantum vacuum. While neutrinos must stay confined inside the horizon, the antineutrinos should be violently ejected. Hence, a black hole (made from matter) should behave as a point-like source of antineutrinos. Our simplified calculations suggest, that the antineutrinos emitted by supermassive black holes in the centre of the Milky Way and Andromeda Galaxy, could be detected by the new generation of neutrino telescopes.

## 1. Introduction

The gravitational properties of antimatter are still not known. While everyone knows that an apple falls down, no one knows if an "anti-apple" would fall down or up. The answer on this question may come from the AEGIS experiment [1] at CERN, designed to compare the gravitational acceleration of atoms of hydrogen and antihydrogen. In this Letter we present an intriguing possibility: if antihydrogen falls up, the supermassive black holes should be emitters of antineutrinos, what may be observable with the new generation of neutrino telescopes (Ice Cube and KM3NeT).

Let us consider a hypothetical gravitational repulsion between matter and antimatter ("antigravity") defined through relations:

$$m_i = m_g \; ; \; m_i = \overline{m}_i \; ; \; m_g + \overline{m}_g = 0 \qquad (1)$$

Here, a symbol with a bar denotes antiparticles; while indices $i$ and $g$ refer to the inertial and gravitational mass (gravitational charge). The first two relations in (1) are experimental evidence [2, 3], while the third one is the conjecture of antigravity which dramatically differs from the mainstream conviction $m_g - \overline{m}_g = 0$, implying (together with the Newton law of gravity) that matter and antimatter are mutually repulsive but self-attractive. Of course, our main premise $m_g + \overline{m}_g = 0$ should be considered as a testable scientific speculation, not excluded by the existing experimental and observational evidence.

At first sight, it may be concluded, that in our Universe, apparently dominated by matter, the gravitational properties of antimatter are not important; the miniscule quantities of antimatter could not have any significant impact. However this naive point of view neglects the physical vacuum, in which, according to Quantum Field Theory, virtual matter and antimatter "appear" in equal quantities. Hence, the gravitational mass of the quantum vacuum (and thus the fate of the Universe) depends on the gravitational properties of antimatter.

Three major consequences of the conjecture (1) are:
1. *A virtual particle-antiparticle pair is a system with zero gravitational mass.*
2. *A virtual pair may be considered as a virtual gravitational dipole.*
3. *A sufficiently strong gravitational field may create particle-antiparticle pairs from the quantum vacuum.*

The idea that antimatter has a negative gravitational charge is not new (For a review see Ref. [4]). What is completely new in our approach is the suggestion that the gravitational properties of antimatter determine the gravitational properties of the quantum vacuum and through the vacuum, antimatter has a major impact in astrophysics and cosmology.

In the present Letter we consider only the third of the above consequences with the particular interest in the creation of neutrino-antineutrino pairs deep inside the horizon of a supermassive black hole. However, it is worth to note, that the most important idea might be to consider the physical vacuum as a fluid of virtual gravitational dipoles and to study the vacuum polarization by an external gravitational field. Our work is in progress to understand if the phenomena, usually attributed to the hypothetical dark matter and dark energy, could be explained as a result of the quantum vacuum polarization by the gravitational field of the known baryonic matter.

As it was demonstrated by Schwinger [5] in the framework of Quantum Electrodynamics, a strong electric field $E$, greater than a critical value $E_{cr}$, can create electron-positron pairs from the quantum vacuum. For instance, electron-positron pairs can be created in the vicinity of an artificial nucleus with more than 173 protons (see for instance Greiner at al. [6] or Ruffini et al. [7]).

In the case of an external (classical i.e. unquantized) constant and homogenous electric field $E$, the exact particle creation rate per unit volume and time is [5, 6]

$$\frac{dN_{m\bar{m}}}{dtdV} = \frac{4}{\pi^2} \frac{c}{\lambdabar_m^4} \left(\frac{g}{g_{cr}}\right)^2 \sum_{n=1}^{\infty} \frac{1}{n^2} \exp\left(-\frac{n\pi}{2} \frac{g_{cr}}{g}\right); \quad g_{cr}(m) = \frac{2c^2}{\lambdabar_m} \qquad (2)$$

where $\lambdabar_m \equiv \hbar/mc$ denotes the reduced Compton wavelength corresponding to the particle with mass $m$. Let us observe that we have replaced the quotient of electric fields $E/E_{cr}$ (appearing in Quantum Electrodynamics) by the quotient of corresponding accelerations $g/g_{cr}$; so that the result (2) could be used not only in the case of an electric field, but also in the case of antigravity.

The Schwinger mechanism is consequence of: (a) the complex structure of the physical vacuum in quantum field theories, and (b) the existence of an external field which attempts to separate particles and antiparticles. In the physical vacuum, short-living "virtual" particle-antiparticle pairs are continuously created and annihilated by quantum fluctuations (which are in fact possible because of Heisenberg uncertainty relation for time and energy). A "virtual" pair can be converted into a real pair only in the presence of a strong external field, which can spatially separate particles and antiparticles, by pushing them in opposite directions; as it does an electric field in the particular case of charged particles. If it is always an attractive force, as commonly believed today, gravity can't separate particles and antiparticles. Hence, the conjectured gravitational repulsion between matter and antimatter is a necessary condition for separation of particles and antiparticles by a gravitational field and consequently for the creation of particle-antiparticle pairs from the quantum vacuum. . But while an electric field can separate only charged particles, gravitation as a universal interaction may create particle-antiparticle pairs of both charged and neutral particles. Thus, "virtual" pairs are spatially separated and converted into real pairs by the expenditure of the external field energy. For this to become possible, the potential energy has to vary by an amount $mg\Delta l > 2mc^2$ in the range of about one Compton wavelength $\Delta l = \hbar/mc$, which leads to the conclusion that the significant pair creation occurs only in a very strong external field $g$, greater than the critical value $g_{cr}$ in Equation (2).

If $g > g_{cr}$, the infinite sum in Eq. (2) has a numerical value not too different from 1. So, a simple, but good approximation is:

$$\frac{dN_{m\bar{m}}}{dtdV} \approx \frac{4}{\pi^2} \frac{c}{\lambdabar_m^4} \left(\frac{g}{g_{cr}(m)}\right)^2 \qquad (3)$$

Of course, the infinite sum in the equation (2) increases very fast with decrease of $g$, and can't be neglected for $g < g_{cr}$.

Sections 2 and 3 in this paper are a preliminary study of the impact of the hypothetical gravitational repulsion between matter and antimatter on the black hole radiation. Of course these rudimentary considerations must be followed with detailed studies, not only by theorists but also by experts in neutrino astronomy. In section 4 we discuss how the hypothesis of antigravity is related with CPT symmetry, general relativity and energy conservation. As an independent hint, that equations (2) and (3) have a meaning for gravitation, an intriguing relation between Hawking temperature and the Schwinger mechanism is demonstrated in the Appendix.

## 2. The Newtonian approximation

For simplicity, let us consider a spherically symmetric gravitational field, created by a spherical body of radius $R_H$ and mass $M$ and let us assume that for all distances $R > R_H$, the gravitational acceleration is determined by the Newton law $g = GM/R^2$. This allows us to define a critical radius $R_{Cm}$ as the distance at which the gravitational acceleration has the critical value $g_{cr}(m)$.

$$R_{Cm} = \sqrt{\frac{GM\bar{\lambda}_m}{2c^2}} \equiv \frac{1}{2}\sqrt{\bar{\lambda}_m R_S} \tag{4}$$

where $R_S = 2GM/c^2$ is the Schwarzschild radius.

Hence, the spherical shell with the inner radius $R_H$ and the outer radius $R_{Cm}$ should be a "factory" for creation of particle-antiparticle pairs with mass $m$. Of course, the creation of particle-antiparticle pairs is also possible outside of this spherical shell (for $R > R_{Cm}$ i.e. $g < g_{cr}$), but according to Eq. (2) it is highly reduced by the exponential factor.

It is evident that there is a series of decreasing critical radii $R_{Cm}$. For instance, according to equation (4), the critical radius $R_{Cv}$ corresponding to neutrinos is nearly four orders of magnitude larger than the critical radius $R_{Ce}$ for electrons, which is about 14 times larger than the critical radius $R_{C\mu}$ for muons, which is 3 times larger than the critical radius $R_{Cn}$ for neutrons and so on. It is obvious that if $R_H > R_{Cm}$, the creation of pairs with mass $m$ is suppressed; for instance, if $R_H = R_{Ce}$, the creation of the more massive pairs, like muon-antimuon and neutron-antineutron is suppressed.

The equation (4) tells us that $R_{Cm} \ll R_S$. Hence, a gravitational field, sufficiently strong to create particle-antiparticle pairs, could exist only deep inside the horizon of a black hole. An immediate consequence is that if (for instance) a black hole is made from ordinary matter, produced particles must stay confined inside the horizon, while antiparticles should be violently ejected because of the gravitational repulsion.

After integration over the volume of the spherical shell, Eq. (3) leads to:

$$\frac{dN_{m\bar{m}}}{dt} \approx \frac{1}{\pi}\left(\frac{R_S}{\bar{\lambda}_m}\right)^2 \frac{c}{R_H} \tag{5}$$

In order to get the result (5) we have assumed $R_H \ll R_{Cm}$, what can be rigorously justified by integration of the exact relation (2) over the interior of the black hole (i.e. from $R_H$ to $R_S$).

According to Eq. (5), the lifetime of a black hole depends on both, mass $M$ and radius $R_H$. If $R_H$ is too small, the black hole is a short living object! For example, if $R_H = R_{Cn}$ (hence the creation of neutron-antineutron pairs is not suppressed), a trivial numerical study, based on the use of the equations (5), reveals that a supermassive black hole would be reduced to one millionth of its initial mass in less than a human life; what is of course in conflict with observations. So, $R_H$ must be larger than $R_{Cn}$.

But a finite $R_H$ demands a mechanism to prevent the collapse. Why the collapse eventually stops for a value $R_H > R_{Cn}$? The answer may be that $R_H = R_{Ce}$ is point of a phase transition, from a neutral to a charged black hole, what would be reflected in the change of the metric from the Schwarzschild metric to the Reissner-Nordstrom metric. To see it, let's imagine that $R_H$ is a little bit smaller than $R_{Ce}$, so that significant creation of electron-positron pairs may start. If electron-positron pairs are produced, an initially neutral black hole must become charged. The negative electric charge of the black hole, opposes the further creation of electron-positron pairs by the gravitational field (it is because the electric and gravitational force on an electron, and also on a positron, have opposite directions); hence, after some time the ejection of positrons should be stopped. As a final result, a black hole made from matter, should emit mainly antineutrinos. The details of the phase transition may be different from this simplified picture, but in order to get a charged black hole (through the mechanism of antigravity) its size must be smaller than the critical radius for electrons, which are the lightest charged particles. Hence, the size $R_H$ of a black hole should satisfy the condition

$$R_{Cn} = \frac{1}{2}\sqrt{\lambdabar_n R_S} < R_H < R_{Ce} = \frac{1}{2}\sqrt{\lambdabar_e R_S} \qquad (6)$$

Now, our main interest is to estimate the number and energy of antineutrinos that should hit the Ice Cube in a certain period of time. In our estimates we will use the upper limit $R_H = R_{Ce}$ in inequality (6), which corresponds to the lower bound for the energy and the number of created neutrino-antineutrino pairs.

Let $d$ and $A_{IC} (\approx 10^6 m^2)$ denote respectively the distance between the source of antineutrinos and the Ice Cube, and the surface of a side of the Ice Cube. Then, according to equations (5) and taking for $R_H$ the upper limit in inequality (6), the number of antineutrinos that may hit the Ice Cube during a time $t$ should be (assuming a negligible absorption/annihilation of antineutrinos on the way to the Ice Cube)

$$N_{\bar{\nu},t} \approx \frac{1}{2\pi^2} \frac{A_{IC}}{d^2} \left(\frac{R_S}{\lambdabar_\nu}\right)^{\frac{3}{2}} \frac{ct}{\sqrt{\lambdabar_\nu \lambdabar_e}} \qquad (7)$$

The energy of the majority of the emitted antineutrinos could be approximated with

$$\varepsilon_{\bar{\nu}} \approx \frac{GM m_\nu}{R_H} = \sqrt{\frac{R_S}{\lambdabar_e}} m_\nu c^2 \qquad (8)$$

This is a consequence of the very fast decrease of creation rate (2) with distance; hence the major fraction of pairs is created at a distance $R$ not too different from $R_H$

Let us look at numerical values. While the mass of neutrino is not precisely known, for the purpose of calculations we use $m_\nu \approx 10^{-37} kg$.

The supermassive black hole in the centre of Andromeda Galaxy is at a distance of about $d_A \approx 770 kpc \approx 2.4 \times 10^{22} m$ and has a mass of $M_A \approx 2.8 \times 10^{38} kg$. Using these values in equations (7) and (8) leads to

$$N_{\bar{\nu}A} \approx 3 \times 10^{10} / year$$
$$\varepsilon_{\bar{\nu}A} \approx 58 GeV \qquad (9)$$

For the supermassive black hole in the centre of the Milky Way, the observed values are $d_{MW} \approx 8 kpc \approx 2.5 \times 10^{20} m$ and $M_{MW} \approx 8.4 \times 10^{36} kg$. Hence

$$N_{\bar{\nu}MW} \approx 1.4 \times 10^{12} / year$$
$$\varepsilon_{\bar{\nu}MW} \approx 10 GeV \qquad (10)$$

Ice Cube detects neutrinos with energies in excess of 100GeV, while its low energy upgrade, dubbed Deep Core [8], decreases the threshold to 10 GeV. Hence, the energies (9) and (10) are just in this interval between 10GeV and 100GeV. Of course, it would be better to have higher energies than our estimates (9) and (10), but even with these energies, the detection at the Ice Cube (and later at KM3NeT) could be possible

The probability to detect an antineutrino may be approximated by

$$P \approx 3.3 \times 10^{-13} E^{2.2} \text{ for } E = 1 - 1000 GeV \qquad (11)$$

what is in fact the formula (9) from Halzen [9]. According to equations (9), (10) and (11) the expected number of detected antineutrinos per year is $\approx 66$ for the supermassive black hole in the centre of the Andromeda Galaxy, and $\approx 108$ for the Milky Way black hole. These numbers are encouraging.

Let us underline once again that because of our choice $R_H = R_{Ce}$, the numerical results (9) and (10) should be a minimum for both, the number of created antineutrinos and their energy. For instance, if $R_H \approx R_{C\mu}$, $\lambdabar_e$ in (7) and (8) should be replaced by $\lambdabar_\mu$; consequently, all numbers in (9) and (10) are larger for one order of magnitude, while the expected number of detected antineutrinos is larger for three orders of magnitude.

The IceCube Neutrino Observatory can discriminate if antineutrinos come from a point-like source or not, but we can't be sure through which mechanism they were created. However, if they exist, antineutrinos produced by antigravity have a few signatures which may help to reveal the mechanism of their creation. We give an incomplete list of signatures noting that each of them should be subject of a detailed study.

First, the energy spectrum of antineutrinos produced by gravity inside the horizon of a black hole should have a sharp peak, approximated by the equation (8).

Second, it is possible to compare energies at the peak, for supermassive black holes in the centre of Andromeda and Milky Way. According to Equation (8)

$$\frac{\varepsilon_{\bar{\nu}A}}{\varepsilon_{\bar{\nu}MW}} = \sqrt{\frac{M_A}{M_{MW}}} \quad (12)$$

i.e. the ratio of the energies at the peak should be equal to the square root of the ratio of masses of black holes, already estimated from independent observations. Let us note that from equations (7) and (11) follows a similar relation for the ratio of the numbers of antineutrinos in the peak.

Third, from the theoretical point of view, the major signature is the strong asymmetry of the antigravity mechanism (only antineutrinos are emitted). However, the Ice Cube has limited possibility to discriminate between neutrinos and antineutrinos (in fact only through Glashow resonance [10], while other detectors, like iron (magnetized) calorimeters, which can distinguish between neutrinos and antineutrinos, suffer from a small size.

### 3. The Schwarzschild black hole

Strictly speaking, the well known Schwarzschild metric

$$ds^2 = c^2\left(1 - \frac{2GM}{c^2 r}\right)dt^2 - \left(1 - \frac{2GM}{c^2 r}\right)^{-1} dr^2 - r^2 d\theta^2 - r^2 \sin^2\theta\, d\phi^2 \quad (13)$$

is the metric „seen" by a test particle attracted by the black hole. The metric „seen" by an antiparticle, which is according to our hypotheses repelled by the black hole, reads

$$ds^2 = c^2\left(1 + \frac{2GM}{c^2 r}\right)dt^2 - \left(1 + \frac{2GM}{c^2 r}\right)^{-1} dr^2 - r^2 d\theta^2 - r^2 \sin^2\theta\, d\phi^2 \quad (14)$$

The major difference is that the metric (13) has, while metric (14) has not a coordinate singularity at $r = 2GM/c^2$.

Of course, while we have given it a physical interpretation, the negative mass Schwarzschild solution (14) is not our invention. It is known for a long time (see for instance Preti and de Felice [11]; Luongo and Quevedo [12]) but considered as unphysical. It serves as the simplest example of a naked singularity and a repulsive space-time allowed by mathematical structure of general relativity but rejected as unphysical.

According to the metric (14) the radial motion of a massive antiparticle (see any good textbook on general relativity, for instance [29]) is determined by

$$\dot{r}^2 = c^2(k^2 - 1) - \frac{2GM}{r} \quad (15)$$

where $k$ is a constant of motion and dot indicates derivative with respect to the proper time $\tau$.

Differentiating the equation (15) with respect to $\tau$ and dividing through $\dot{r}$ gives

$$\ddot{r} = \frac{GM}{r^2} \qquad (16)$$

which has the same form as should have the corresponding Newtonian equation of motion with the assumed gravitational repulsion.

Additionally, for an antiparticle dropped from rest at $r = R$ (we are in particular interested in the case $R \ll R_S$), the equation (15) can be written as

$$\dot{r}^2 = 2GM\left(\frac{1}{R} - \frac{1}{r}\right) \qquad (17)$$

This has the same form as should have the Newtonian formula equating the gain in kinetic energy to the loss in gravitational potential energy for an antiparticle of unit mass falling from rest at $r = R$.

Of course, in spite of the same form of Newtonian equations and equations (16) and (17) coming from general relativity, there are fundamental differences between them. The coordinate $r$ in equations (16) and (17) is not the radial distance as it is in Newtonian theory, and dots indicate derivatives with respect to proper time $\tau$ rather than universal time $t$. However the same form of equations is a justification of our work in the previous section; but we must be aware of the different meaning of some quantities; for instance the quantity $R_H$ defined in the previous section should not be considered as the radius of a sphere but rather as the Schwarzschild coordinate $r$ of the surface of the black hole.

## 4. Antigravity, CPT symmetry, general relativity, and energy conservation

The idea of antigravity has appeared soon after the discovery of antiparticles. However, in the early sixties of the 20[th] century, it was suppressed, not by experimental evidence but by theoretical arguments [13-16] which seemed very strong at that time. In a critical review at the end of the last century [4], these classical arguments are still considered as sufficient to exclude antigravity, but some weak points of arguments were pointed out as well. Schiff's argument [14, 15] suffers from incorrect renormalization procedure, while Good's argument [16] is criticised because of a critical dependence on the use of an absolute gravitational potential. Hence, the Morrison's argument [13](in the form of a gedanken experiment pointing out an energy non-conservation paradox) has remained the most compelling theoretical objection against antigravity. The arguments against antigravity were further questioned by Chardin and Rax [17-19]).

Independently of the debate concerning antigravity arguments, a scenario based on a tiny Lorentz and CPT violation in the Standard-Model Extension allows a difference of 50% between gravitational acceleration of matter and antimatter [20-21].

In a way, the 20[th] century was spent in theoretical efforts to understand if antigravity is possible or not and in thinking about and preparing for the future experiments. Theoretically the question is still open, while the crucial experiments like AEGIS are on the way. While our approach is to postulate antigravity and study the consequences, in the present section we briefly address how antigravity is related to CPT symmetry, general relativity and energy conservation.

Apparently, the antigravity is compatible with the quantum field theory. Firstly, as well known, antigravity does not violate the CPT symmetry [4]. Popularly speaking, CPT only implies that an anti-apple falls toward an anti-earth in the same way as an apple falls toward an earth; CPT says nothing how an anti-apple falls toward an earth. But let us note that in a recent seminal publication [22], for the first time it was argued that CPT symmetry might favour antigravity. Secondly, antigravity is consistent with our current wisdom that gravity is mediated by graviton, i.e. a tensor (spin-two) particle. In a gravity theory allowing for positive and negative gravitational charge and mediated by a tensor particle, like charges attract, while unlike charges repel each other [23, 24]).

Of course the hypothesis of antigravity violates the weak equivalence principle (i.e. the equivalence of inertial and gravitational mass) which is the cornerstone of general relativity. However this very particular kind of violation might allow preserving the geometry of general relativity. For instance, a particle and an antiparticle dropped from rest at $r = R$, of a Schwarzschild black hole with metric (13), should follow the same timelike geodesic in the opposite directions.

Concerning the violation of conservation of energy in different variants of the Morrison's gedanken experiment [4, 13, 25] the question is reopened by interesting arguments [17-19] refuting the idea that the Morrison gedanken experiment forbids antigravity.

While the Morrison experiment is correctly described by many authors, it is surprising that none of them pays attention to the Appendix of the original paper (Morrison [13] pages 367-368) where Morrison discusses necessary physical assumptions to prevent the perpetual motion machine; hence the outcome of his gedanken experiment is model dependent.

It should be added that there is one common mistake in all post-Morrison papers. Everyone assumes that a particle-antiparticle system like positronium is not affected by gravity at all; so the position of the system can be changed without work, what is the reason for the violation of conservation of energy. However, positronium should be considered as a gravitational dipole and its radial distance from the centre of the Earth can't be changed without work.

## 5. Comments

Of course, the black hole radiation considered in this Letter is quite different and much stronger than the famous Hawking radiation. It may be just one of many phenomena that could be eventually related to the hypothesis of the gravitational repulsion between matter and antimatter. As suggested in a number of previous publications antigravity might be an alternative to the inflation in cosmology [26]) and ground for the understanding of the cosmological constant problem and dark energy [27], matter-antimatter asymmetry in the Universe [26] and the Pioneer anomaly [28].

Let us note that non-thermal radiation of black holes may exist even if there is no antigravity. To see it let imagine that there is a still undiscovered interaction, repulsion between matter and antimatter, with the following features: (a) The new force acts, between particles having an appropriate "charge", which (just as the electric charge) can be positive or negative. In order to be definite, a positive "charge" is attributed to matter and a negative one to antimatter. (b) There is an attractive force between "charges" of the same sign, and a repulsive force between "charges" of different sign. This is just opposite to the familiar case of electric charges. (c) The assumed repulsion between particles and antiparticles is stronger than the gravitational attraction between them. (d) It may be a force with much shorter range of interaction than gravitation. Such a new force is less elegant possibility than antigravity but it is evident that it can produce particle-antiparticle pairs according to equation (2).

If eventually a non-thermal radiation of black holes is detected at the Ice Cube, it would be impossible to know if it is result of antigravity or a new force described above. In such a case comparison with laboratory experiments would be crucial. For instance if there is a non-thermal radiation off black holes, but antigravity is excluded by the AEGIS experiment, it would be signature of a new force.

## Appendix: The Hawking temperature from Schwinger's formula

It is intriguing that the famous expression for the Hawking temperature

$$k_B T = \frac{\hbar c^3}{8\pi GM} \qquad (18)$$

can be accurately estimated from the Schwinger formula (2).

The equation (2) trivially transforms to

$$\frac{dN_{m\bar{m}}}{dtdV} = \frac{1}{\pi^2} \frac{g^2}{c^3} \frac{1}{\lambdabar_m^2} \sum_{n=1}^{\infty} \frac{1}{n^2} \exp\left(-n\pi \frac{c^2}{g\lambdabar_m}\right) \qquad (19)$$

Taking only the leading term $n=1$, distribution (19) has a maximum for

$$\lambdabar_{max} = \frac{\pi}{2} \frac{c^2}{g} \qquad (20)$$

The equation (20) together with the Wien displacement law $\lambda_{max} T = b$ (where $T$ and $b$ are respectively the absolute temperature and Wien displacement law constant) leads to

$$k_B T = A \frac{1}{2\pi} \frac{\hbar}{c} g \qquad (21)$$

where $k_B$ is the Boltzmann constant and $A$ a dimensionless constant

$$A = \frac{2}{\pi} \frac{b k_B}{\hbar c} \approx 0.8 \qquad (22)$$

Thus, starting with the Schwinger mechanism we have attributed a temperature $T$ to the vacuum around a massive body. In the particular case $g = GM/R_S^2$ the equation (21) transforms to

$$k_B T = A \frac{\hbar c^3}{8\pi GM} \qquad (23)$$

what is close to the exact result (18) corresponding to the value $A = 1$.